\documentclass[12pt]{article}
\usepackage{array}
\usepackage{graphicx}
\usepackage{amssymb}
\usepackage{amsmath}
\usepackage{multirow}
\input{epsf}
\usepackage{cite}
\def\@fmsl@sh#1#2#3{\m@th\ooalign{$\hfil#1\mkern#2/\hfil$\crcr$#1#3$}}
 \def\eq#1\en{\begin{equation}#1\end{equation}}
\def\s[#1,#2]{[#1\stackrel{\star}{,}#2]}
\def\sx[#1,#2]{[#1\stackrel{\star_{x}}{,}#2]}

\newcommand{\nc}{\newcommand}
\nc{\beq}{\begin{equation}}
\nc{\eeq}{\end{equation}}
\nc{\beqa}{\begin{eqnarray}}
\nc{\eeqa}{\end{eqnarray}}

\def\bc{\begin{center}}
\def\ec{\end{center}}

\def\to{\rightarrow}

\def\gsim{\mathrel{\mathpalette\atversim>}}

\def\bc{\begin{center}}
\def\ec{\end{center}}

\def\gsim{\mathrel{\rlap{\lower4pt\hbox{\hskip1pt$\sim$}}

    \raise1pt\hbox{$>$}}}       

\def\gsim{\mathrel{\rlap{\lower4pt\hbox{\hskip1pt$\sim$}}
    \raise1pt\hbox{$>$}}}       

\begin{document}
\makeatletter
\def\fmslash{\@ifnextchar[{\fmsl@sh}{\fmsl@sh[0mu]}}
\def\fmsl@sh[#1]#2{%
  \mathchoice
    {\@fmsl@sh\displaystyle{#1}{#2}}%
    {\@fmsl@sh\textstyle{#1}{#2}}%
    {\@fmsl@sh\scriptstyle{#1}{#2}}%
    {\@fmsl@sh\scriptscriptstyle{#1}{#2}}}
\def\@fmsl@sh#1#2#3{\m@th\ooalign{$\hfil#1\mkern#2/\hfil$\crcr$#1#3$}}
\makeatother

\thispagestyle{empty}
\begin{titlepage}
\boldmath
\begin{center}
  \Large {\bf The flavor of quantum gravity}
    \end{center}
\unboldmath
\vspace{0.2cm}
\begin{center}
{ {\large Xavier Calmet}\footnote{x.calmet@sussex.ac.uk},
{\large Dionysios Fragkakis} \footnote{d.fragkakis@sussex.ac.uk}
 and
  {\large Nina Gausmann} \footnote{n.gausmann@sussex.ac.uk}
}
 \end{center}
\begin{center}
{\sl Physics and Astronomy, 
University of Sussex,   Falmer, Brighton, BN1 9QH, UK 
}
\end{center}
\vspace{\fill}
\begin{abstract}
\noindent
We develop an effective field theory to describe the coupling of non-thermal quantum black holes to particles such as those of the Standard Model. The effective Lagrangian is determined by imposing that the production cross section of a non-thermal quantum black hole be given by the usual geometrical cross section. Having determined the effective Lagrangian, we estimate the contribution of a virtual hole to the anomalous magnetic moment of the muon, $\mu \to e \gamma$ transition and to the electric dipole moment of the neutron. We obtain surprisingly weak bounds on the Planck mass due to a chiral suppression factor in the calculated low energy observables. The tightest bounds come from $\mu \to e \gamma$ and the limit on the neutron electric dipole moment. These bounds are in the few TeV region. However, the bound obtained from proton decay is much more severe and of the order of $1 \times 10^6$ GeV. 
\end{abstract}  
\end{titlepage}



\newpage

Collider physics is obviously a powerful tool to produce new particles and hence to probe new physics. However, it is often useful to use precise low energy measurements to search for new physics effects. In particular if the new physics model leads to new sources of violation of flavor conservation, the effects in rare decays can be dramatic. New physics effects can be even larger if CP is violated. The aim of this work is to consider flavor physics within the realm of non-thermal small black holes  in a model independent way as far as possible. Our  goal is to  bound the energy scale at which quantum gravity effects become important, i.e. the Planck scale, using data on the anomalous magnetic moment of the muon,  low energy experiments searching for violation of lepton flavor conservation and bounds on a neutron electric dipole moment. 

The value of the Planck scale is very poorly known. One of the major theoretical developments of the last 15 years has been the realization that the reduced Planck scale could be anywhere between a few TeVs and 10$^{18}$ GeV if there are large extra-dimensions \cite{ArkaniHamed:1998rs,Randall:1999ee} or a large hidden sector \cite{Calmet:2008tn,Calmet:2010nt}. There are several experimental  constraints on these theoretical frameworks  leading to complete exclusion of the model to bounds in the 1 to 100 TeV region  depending on the number of extra-dimensions.  One of the most remarkable effects of quantum gravity physics is without a doubt the formation of small black holes in the collision of particles. It is now well appreciated that semi-classical black holes cannot be produced at the Large Hadron Collider because the center of mass energy is not sufficient even if the Planck scale is around a few TeV. However, quantum black holes, i.e., non-thermal small black holes with masses of the order of the Planck scale could be produced abundantly \cite{Meade:2007sz,Calmet:2008dg}.  The first experimental papers setting limits on the masses of these holes have started to appear, see e.g. \cite{Aad:2011aj}.

A black hole is characterized by its mass, spin and electric charge. From a particle physicist's point of view, however, the most important characteristic of a black hole is its production cross section. This property is the one which will fix the effective theory in an unambiguous way.
 Non-thermal quantum black holes can be thought of as states which are created and decay almost instantly, we will thus treat them as short-lived gravitational  states. We can model these states using quantum fields and their interactions using the language of effective field theories. This is natural since these holes only couple to a few particles. Our strategy is the following; we will model a spinless non-thermal quantum black hole using a scalar field which can be charged under the gauge quantum numbers of the Standard Model, but it could also be neutral. The interactions of this scalar field with the particles of the Standard Model will be chosen in such a way that the geometrical cross section $\sigma(\mbox{particle 1 + particle 2 $\to$ QBH})= \pi r_s^2$, where $r_s$ is the Schwarzschild radius, is obtained when calculating $\sigma(\mbox{particle 1 + particle 2 $\to$ $\phi$})$ where the scalar field $\phi$ represents the spinless quantum black hole. Although we will focus on the four dimensional case and on spinless quantum black holes, our results can be easily generalized to the case of extra-dimensional black holes and to the case of black holes with spin.

Very little is known about quantum gravitational physics. However, any consistent theory of quantum gravity should preserve the gauge symmetries of the Standard Model of particle physics. This implies that gauge quantum numbers must be preserved by quantum gravitational interactions including non-thermal quantum black holes. On the other hand, global symmetries can be violated by quantum gravitational interactions. Examples of Lorentz violating vacua are known in string theory \cite{Kostelecky:1988zi}. Lorentz violation effects typically lead to very tight bounds on the scale of quantum gravity. Here we shall assume that Lorentz is not violated and focus on violation of flavor symmetry and CP violation induced by quantum black hole processes. 

Before we start discussing the interactions of quantum black holes with elementary particles of the Standard Model, let us discuss our assumptions concerning the holes themselves. A black hole can be uniquely determined by its mass, spin and charge. From that point of view, the idea of describing a black hole by a massive quantum field carrying a spin and a charge is reasonable. However, there are some subtleties in the case of quantum black holes. First of all, if they were created in  collisions of quarks or gluons, one would expect them to carry a QCD color charge as well since this gauge quantum number cannot be violated. Furthermore, it is not obvious whether their mass spectrum is continuous or discrete. We posit the following Lagrangian for a spinless and neutral quantum black hole $\phi$:
\begin{eqnarray}
  L_4 = \frac{1}{2} \partial_\mu \phi \partial^\mu \phi -\frac{1}{2}M_{BH}^2 \phi^2.
  \end{eqnarray}
One expects the first quantum black mass to be of the order of the Planck mass. If the spectrum is discrete, one can consider a collection of scalar fields. On the other hand, if the mass spectrum was continuous, the mass of the quantum black hole would increase with the energy of the process. Note that the latter case resembles very much an unparticle field. In the following we assume a discrete mass spectrum and focus on the lightest quantum black hole, but this will not impact our results which could be trivially extended to describe a continuous mass spectrum.

\begin{figure}[ht]
\begin{center}
  \includegraphics[width=50mm]{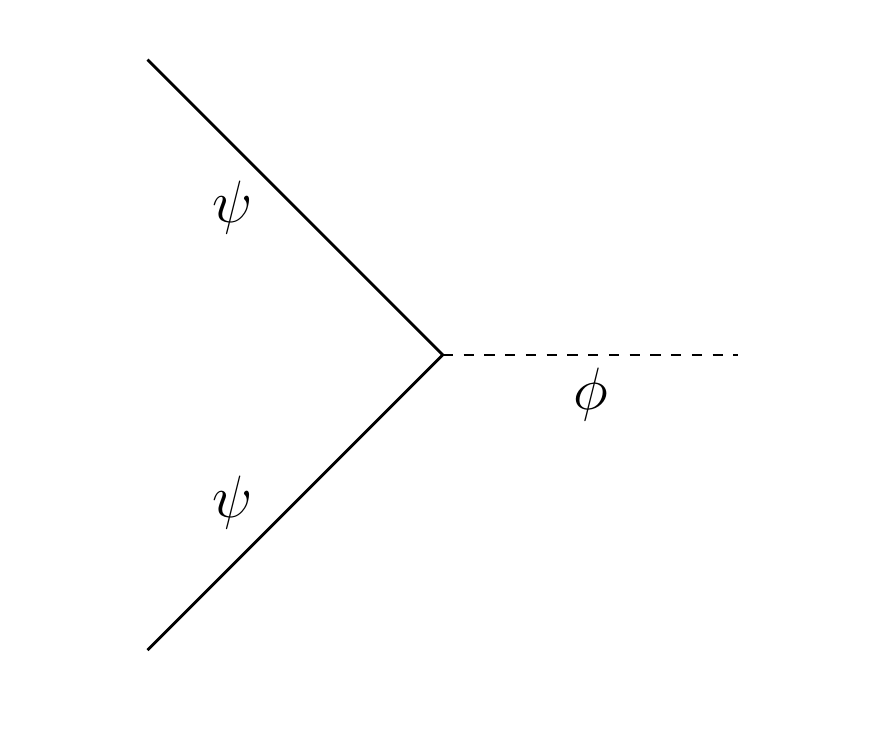}
  \caption{Vertex generated by the effective Lagrangian (\ref{effLagr}) which describes the annihilation of two fermions into a quantum black hole.} 
\end{center}
\end{figure}

Let us start from the effective Lagrangian
\begin{eqnarray} \label{effLagr}
  L_6 = \frac{c}{\bar{M}_{P}^{\,2}} \, \partial_\mu \partial^\mu \phi \bar{\psi}_1 \psi_2 + h.c.
  \end{eqnarray}
where $c$ is a parameter which will be adjusted to yield the correct cross section, $\bar M_P$ is the reduced Planck mass, $\phi$ is a scalar field describing the non-thermal quantum black hole, $\psi_i$ is a fermion field which could be any fermion of the Standard Model. We shall assume here that $\phi$ is neutral. The gauge charges of $\psi_1$ must thus be matched by those of $\psi_2$. Once again one can easily extend this method to charged quantum black holes. Including a spin would require introducing fields with  spins which can be done without any difficulty as well. We start with dimension 6 operators since in 4-dimensions, the cross section for black hole production goes as $M_P^{-4}$.
One obtains the following cross section for $\phi$ production:
 \begin{eqnarray}
   \sigma \left( 2 \psi \rightarrow \phi \right) &=& \frac{\pi}{s} |\mathcal{A}|^{2} \, \delta( s - M_{BH}^{2})
  \end{eqnarray}
  where $M_{BH}$ is the mass of the $\phi$ field, $s=(p_1+p_2)^2$,  $p_1$ and $p_2$ are respectively the four-momenta of $\psi_1$ and $\psi_2$. One finds  
   \begin{eqnarray}
   |\mathcal{A}|^{2} &=& s^{2} \frac{c^{2}}{\bar{M}_{P}^{\,4}} \left[ s - \left( m_{1} + m_{2} \right)^{2} \right],
  \end{eqnarray}
where $m_1$ and $m_2$ are the masses of the fermions $\psi_1$ and $\psi_2$. We can now compare this cross section to the geometrical cross section. We assume that quantum black holes have the same cross section as semi-classical ones:
  \begin{eqnarray}
   \sigma &=& \pi r_{s}^{2}
  \end{eqnarray}
where the four-dimensional Schwarzschild radius is given by
  \begin{eqnarray}
   r_s \left( s , 0 , \bar{M}_{P} \right) &=& \frac{\sqrt{s}}{4 \pi \bar{M}_{P}^2}
  \end{eqnarray}
and one thus finds:
 \begin{eqnarray} \label{c}
   c^{2} = \frac{1}{4 \pi \left[ (p_1+p_2)^2 - \left( m_{1} + m_{2} \right)^{2} \right]}\frac{\sqrt{(p_1+p_2)^2} \left[ \left(\sqrt{(p_1+p_2)^2} - M_{BH}\right)^{2} + \frac{\Gamma^{2}}{4} \right]}{\Gamma}
  \end{eqnarray}
where $\Gamma$ is the decay width of $\phi$. Note that we have used the representation
 \begin{eqnarray}
   \delta( s - M_{BH}^{2})    &=& \frac{\Gamma}{4 \pi \sqrt{s} \left[ \left(\sqrt{s} - M_{BH}\right)^{2} + \frac{\Gamma^{2}}{4} \right]}
    \end{eqnarray}
for the delta function. The partial width of $\phi$ can be estimated as follows:
 \begin{eqnarray}
   \Gamma\left(QBH \rightarrow \psi_1 + \psi_2 \right) 
   &\sim& \frac{M_{BH}}{64 \pi^{2}},
  \end{eqnarray}
since there are about 100 degrees of freedom in the Standard Model, the total width is about hundred times larger. We can use this relation to estimate $\Gamma$ in eq. (\ref{c}). 

The dimension 6 operators introduced in (\ref{effLagr}) have interesting consequences for precise low energy measurements such as the anomalous magnetic moment of the muon, rare decays forbidden in the Standard Model or they could even lead to new sources of CP violation if a $\gamma_5$ is introduced in the effective Lagrangian describing the coupling of the quantum black hole to the fermions
\begin{eqnarray} \label{effLagr2}
  L_{6, CP} = \frac{c}{\bar{M}_{P}^{\,2}} \, \partial_\mu \partial^\mu \phi \bar{\psi}_1 i \gamma_5  \psi_2 + h.c.
  \end{eqnarray}
 It is easy to estimate the one loop induced effective Lagrangian in the low energy regime. Obviously the effective theory is not renormalizable, but we can use power counting arguments. We find 
  \begin{eqnarray}
L_{eff} = \frac{e}{2} \frac{1}{16 \pi^2} \sum_{i j}  \frac{m_i}{\bar{M}_P^2}  \bar{\psi_i} (A_{ij}+ B_{ij} \gamma_5) \sigma_{\mu\nu} \psi_j F^{\mu\nu} 
  \end{eqnarray}
where $m_i$ is the mass of the heaviest of the two fermions, $A_{ij} = A$ and $B_{ij} = B$ are numerical coefficients which are found to be of order 1. We took the momentum cutoff of the loop integral to be of the order of the reduced Planck mass. Note that we have carefully considered the Dirac structure of the loop diagram. This led to the chiral suppression factor $m_i/\bar M_P$.

Let us first consider the contribution to the magnetic moment of the muon. One finds the well known result
\begin{eqnarray}
\Delta a = \frac{1}{16 \pi^2}\frac{ m^2_\mu}{\bar M^2_P} A.
\end{eqnarray}
Using $\Delta a \sim 10^{-9}$  \cite{pdg} we obtain:
\begin{eqnarray}
\bar{M}_{P} &>& 266 \, \mathrm{GeV}.
  \end{eqnarray}
  Note that this  bound is obtained with very little assumptions, we only assumed that quantum black holes can be treated as virtual objects which couple to low energy modes. It is surprisingly weak and indicates that quantum black holes could very well be relevant for LHC physics.
  \begin{figure}[ht]
  \begin{center}
  \includegraphics[width=70mm]{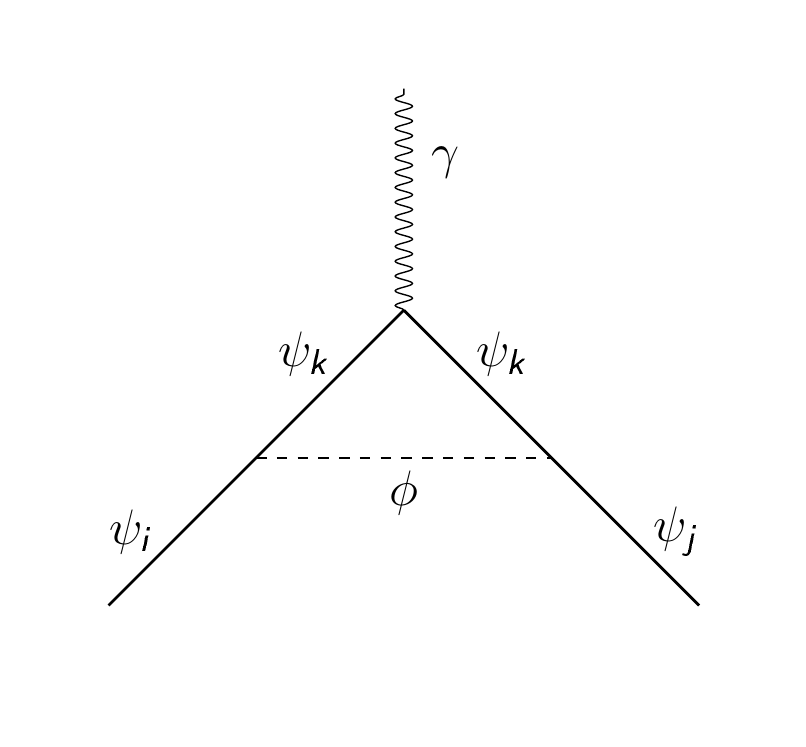}
  \label{}
  \caption{Quantum black hole contribution to the anomalous magnetic moment of the muon, rare decays of leptons or the EDM of the fermions of the Standard Model.}
\end{center}
\end{figure}

Bounds on lepton flavor violating processes also lead to limits on the reduced Planck mass. Unless  the lepton number is gauged, quantum black hole processes are expected to lead to transitions which do not preserve lepton number. We find
 \begin{eqnarray}
   \Gamma( \mu \rightarrow e\gamma ) 
   &=& e^2   \frac{A^{2}}{1024 \pi^5} \frac{m_{\mu}^{5}}{\bar{M}_{P}^{\,4}}
     \\
   \Gamma( \tau \rightarrow e\gamma ) 
   = \Gamma( \tau \rightarrow \mu\gamma ) 
   &=& e^2 \frac{A^{2}}{1024 \pi^5} \frac{m_{\tau}^{5}}{\bar{M}_{P}^{\,4}}. 
  \end{eqnarray}
The current experimental limit for the muon decay into an electron and photon is $\mathrm{Br}( \mu \rightarrow e\gamma ) < 1.2 \times 10^{-11}$ and accordingly for the tau $\mathrm{Br}( \tau \rightarrow \mu \gamma ) < 4.4 \times 10^{-8}$  \cite{pdg}. We find the following limit on the reduced Planck mass 
  \begin{eqnarray}
  \bar{M}_{P} &>& 3  \times 10^{4} \, \mathrm{GeV}
   \end{eqnarray}
  using the bound on the transition $\mu \to e \gamma$ and 
  \begin{eqnarray}
\bar{M}_{P} &>& 3  \times 10^{3}  \, \mathrm{GeV}
  \end{eqnarray}
  using the bound on the transition $\tau \to \mu \gamma$.
  
  If CP is violated by quantum black hole processes, the effective Lagrangian also gives a contribution to the electric dipole moment of leptons and quarks of the Standard Model. 
The Lagrangian yields  the following electric dipole moment of the electron
  \begin{equation}
   d(e) =  \frac{e  B}{16 \pi^2} \frac{m_e}{\bar{M}^2_{P}}.
  \end{equation}
Using the current experimental constraint  $d(e) = (0.07 \pm 0.07) \times 10^{-26} \mathrm{e \, cm}$, we find a bound on the reduced Planck mass:
\begin{eqnarray}
\bar{M}_{P} &>&  1 \times 10^{4} \, \mathrm{GeV}
  \end{eqnarray}
while using the bound for the muon, i.e. $d(\mu) = (-0.1 \pm 0.9) \times 10^{-19} \mathrm{e \, cm}$, we find 
\begin{eqnarray}
\bar{M}_{P} &>& 36  \, \mathrm{GeV}.
  \end{eqnarray}
  
Finally, there is also a contribution to the electric dipole moment of the neutron. The current bound on the neutron electric dipole moment is $d(n) = 0.29 \times 10^{-25} \mathrm{e \, cm}$ \cite{pdg}. One finds
  \begin{equation}
   d(n) = \frac{4}{3} d(d) - \frac{1}{3} d(u) = \frac{ e B}{16 \pi^2 \bar{M}_{P}} \left( \frac{4}{3} \frac{m_d}{\bar{M}_{P}} - \frac{1}{3} \frac{m_u}{\bar{M}_{P}} \right)
  \end{equation}
and a bound on the reduced Planck mass of
  \begin{eqnarray}
   \bar{M}_{P} \approx  5 \times 10^{3} \, \mathrm{GeV}
  \end{eqnarray}
where we took $B\sim 1$ which followed from our estimate. 

Although this is not the main purpose of our article, our result can also be used in the context of collider physics. For processes at energies well below the first quantum back hole mass, one  can integrate out the fields $\phi$ and finds dimension six operators:
   \begin{eqnarray}
   \mathcal{L} &=&\frac{c^{2}}{\bar{M}_{P}^{\,4}} \frac{s^2}{M_{BH}^{\,2}}  \bar{\psi} \psi \bar{\psi} \psi
  \end{eqnarray}
for $M_{BH}^{\,2} \gg s$. These operators can be used to describe non-thermal quantum black hole production in the collisions of quarks at collider. Our result can easily be generalized to include (anti-) quark + gluon or gluon  + gluon processes. These processes can easily be implemented into Madgraph or Calchep.

Dimension 6 operators  similar to those discussed above can be generated by the exchange of a quantum black hole field which violates the baryon and lepton numbers. These operators mediate proton decay. For the lifetime of the proton we find:
   \begin{eqnarray}
 \tau = \frac{\bar{M}_{P}^{\,12}}{m_p^5 c^4 \Lambda_{QCD}^8} \sim \frac{10^4}{( 16 \pi)^2} \frac{\bar{M}_{P}^{\,10}}{m_p^5\Lambda_{QCD}^6}
   \end{eqnarray}
where $\Lambda_{QCD}$ is the typical energy of the quarks in the proton. Using  $\tau> 5 \times 10^{33} y$ \cite{pdg}, we find $M_P> 1 \times 10^6$ GeV. If low scale quantum gravity is the solution to the weak hierarchy problem, the proton decay problem must be addressed. An obvious solution is to gauge the baryon or lepton number.

Using an effective Lagrangian formulation for non-thermal quantum black holes we have derived  bounds on the reduced Planck mass. The reasonably good agreement between the experimental value of the anomalous magnetic moment of the muon  and the theoretical predictions leads to a bound on the Planck mass of the order of $266$ GeV. If discrete symmetries such as lepton number or CP are violated by quantum gravitational physics, one obtains even tighter limits on the energy scale at which quantum gravity effects become sizable, respectively $10^{4}$ GeV and $10^{3}$ GeV. These bounds are surprisingly weak.  The limit coming from the lack of observation of proton decay is much tighter and unless this decomposition is  forbidden by some mechanism, it represents the tightest limit to date on the reduced Planck mass. Note that cosmic rays allow one to set a limit on the four dimensional Planck mass of about 500 GeV \cite{Calmet:2008rv}, but there are several theoretical uncertainties in that bound. In contrast, the limits presented in this paper are rather model independent.
A consequence of our result is that the assumption made in \cite{Calmet:2008dg} concerning the coupling of quantum black holes to long wave modes or their lack of virtuality can be relaxed. Unless quantum gravity violates CP or baryon number, quantum black holes could well be within the reach of the LHC.

\bigskip

\bigskip
\bigskip


\end{document}